\documentclass[referee]{raa}           
\usepackage{graphicx,times}
\usepackage{natbib}
\usepackage{amssymb,amsmath}
\usepackage{multirow}
\bibpunct{(}{)}{;}{a}{}{,}

\usepackage[a4paper=true,dvipdfm=true,pagebackref=true]{hyperref}
\hypersetup{pdftitle = The title of my PDF, pdfauthor = My name, pdfsubject= The subject, pdfkeywords = keyword1 keyword2 keyword3} 
\hypersetup{colorlinks = true, linkcolor = green, anchorcolor = red, citecolor = blue, filecolor = red, pagecolor = red, urlcolor = red}

\begin{document}

   \title{System solutions study on the fatigue of the fast cable-net structure caused by form-changing operation
$^*$
\footnotetext{\small $*$ Supported by the Young Scientist Project of the Natural Science Foundation of China (Grant No. 11303059) and the Chinese Academy of Sciences Youth Innovation Promotion Association.}
}

 \volnopage{ {\bf 2012} Vol.\ {\bf X} No. {\bf XX}, 000--000}
   \setcounter{page}{1}

   \author{Peng Jiang\inst{1,2}, Ren-Dong Nan\inst{1,2}, Lei Qian\inst{1,2}, You-Ling Yue\inst{1,2}}

   \institute{ National Astronomical Observatories, Chinese Academy of Sciences, Beijing 100012, China; {\it pjiang@bao.ac.cn}\\
        \and
             Key Laboratory of Radio Astronomy, Beijing 100012, China\\
\vs \no
   {\small Received 2012 June 12; accepted 2012 July 27}
}

\abstract{The Five-hundred-meter Aperture Spherical Radio Telescope (FAST) is supported by a cable-net structure, whose change in form leads to a stress range of approximately 500MPa. This stress range is more than twice the standard authorized value. The cable-net structure is thus the most critical and fragile part of the FAST reflector system. In this study, we first search for a more appropriate deformation strategy that reduces the stress amplitude generated by the form-changing operation. Second, we roughly estimate the tracking trajectory of the telescope during its service life, and conduct an extensive numerical investigation to assess the fatigue resistance requirements. Finally, we develop a new type of steel cable system that meets that cable requirements for FAST construction. 
\keywords{five-hundred-meter aperture spherical radio telescope, fatigue resistance, astronomical techniques and approaches, cable-net structure, finite element
}
}

   \authorrunning{P. Jiang et al. }            %author_head in even pages
   \titlerunning{System solutions study on the fatigue}  % title_head in odd pages
   \maketitle

%________________________________________________ sections below
%
\section{Introduction}           %% first-level sections will be auto-capitalized
\label{sect:intro}

The Five-hundred-meter Aperture Spherical Radio Telescope (abbreviated as FAST), one of the key scientific projects of the national 11th Five-year Plan, was approved for construction by the National Development and Reform Commission on July 10, 2007. FAST will make observations at frequencies from 70MHz to 3GHz. The design resolution and pointing accuracy will be $2.9'$ and $8''$ respectively. To achieve these technical specifications, the root-mean-square value of the out-of-plane error of the reflector should be no more than 5$mm$ (\citealt{1,2}).

\begin{figure}[!htm]
\centering
\includegraphics[width=0.5\textwidth]{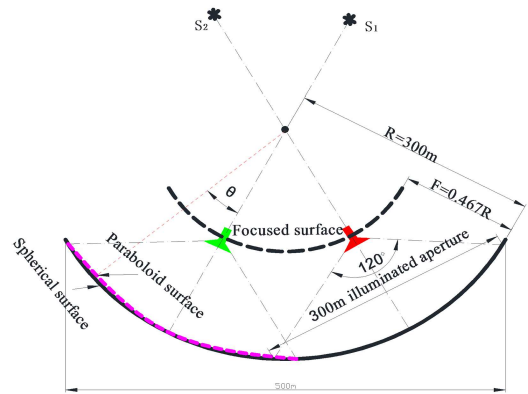}
\caption{Geometric optical principle of FAST}
\label{fig:1}
\end{figure}

\begin{figure}[!htm]
\centering
\includegraphics[width=0.5\textwidth]{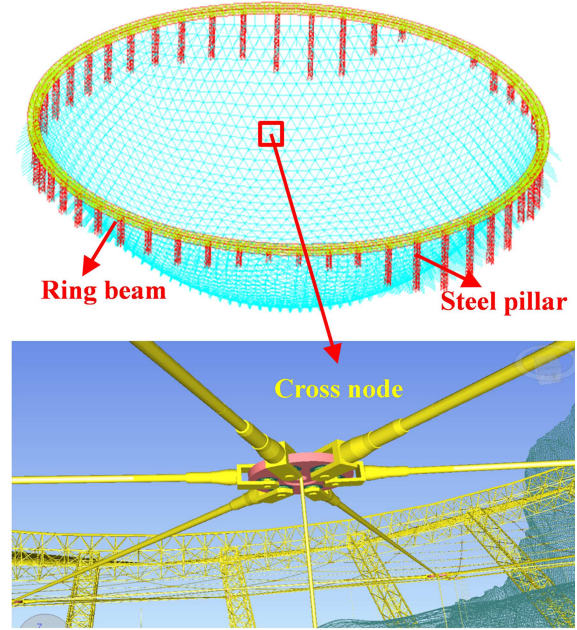}
\caption{Concept of the adaptive cable-net structure}
\label{fig:2}
\end{figure}

According to the geometric optical principle of FAST (illustrated in Figure \ref{fig:1}), the supporting structure of the reflector system should be capable of forming a parabolic surface from a spherical surface. This is the most prominent special requirement of the telescope beyond those of conventional structures. The National Astronomy Observatory of China has been carrying out a rigorous feasibility study of critical technologies since 1994. More than 100 scientists and engineers from 20 institutions are involved in the project. An extensive comparative analysis of several design plans of the supporting structure of the reflector system was performed (\citealt{3,4}). An Arecibo-type plan was selected because a cable-net structure can easily accommodate the complex topography of Karst terrain, thus avoiding heavy civil engineering works between actuators and the ground (see Fig.~\ref{fig:2}) (\citealt{5}).

Later, an extensive numerical comparative analysis was performed among several different cell types, such as three-dimensional cells, Kiewitt cells and geodesic triangular cells. Geodesic triangular cells were selected because their stress distribution is more even (\citealt{5,6}).

The cable net comprises 6670 steel cables and approximately 2225 cross nodes. The lengths of cables range from 10.5 to 12.5$m$, the total weight of the net is 1300 tons, and the cross-sections of main cables have 16 different areas ranging from 280 to 1319$mm^{2}$. None of the cables are
connected, allowing us to set their cross sections according to their loads.

The outer edge of the cable net is suspended from a steel ring beam whose diameter is 500$m$ (see Fig.~\ref{fig:2}). The cross nodes of the cable net are used as control points. Each is connected to an actuator by a down-tied cable. By controlling the actuator using feedback from the measurement and control system, the positions of the cross nodes can be adjusted to form an illuminated aperture having a diameter up to 300$m$. This illuminated aperture moves along the spherical surface according to the zenith angle of the target objects (see S$_{1}$ and S$_{2}$ in Figure \ref{fig:1}).

The above description clearly indicates that long-term observations by FAST will lead to long-term frequent shape-changing operation of the cable-net structure. Previous research has shown that such shape-changing operation introduces stress range on the order of 500$MPa$, which is nearly twice the standard authorized value (\citealt{7,8,9}). The cable-net structure is thus the most critical and fragile part of the FAST reflector system.

To improve the reliability and service life of FAST, the present work, on the basis of the final design, searches for a way to reduce the stress range acting on the cable in shape-changing operation and develops a new type of cable system with ultra high fatigue resistance that meets the requirements of the FAST observation principle.

\section{Optimization of the deformation strategy}

\subsection{Analysis of the main influencing factors}

In general, the illuminated aperture is a parabolic surface, whose profile can be expressed as
\begin{equation}
	x^{2}=2py+c \label{eqn:1}
\end{equation}

The variables and parameters of the problem are the weight of the reflector element $w$, the density of the cable $\rho$, the elastic modulus of the cable $E$, the area of the cable cross section $A_{i}$, the geometric parameters of the illuminated aperture $p$ and $c$, the diameter of the illuminated aperture $d$, the radius of the cable-net structure $R$, and the diameter of the ring beam $D$. The stress amplitude can thus be expressed as
\begin{equation}
\Delta\sigma=f(w, E, \rho, A_{i}, p, c, D, d, R). \label{eqn:2}
\end{equation}

Among these governing parameters, $E$, $d$ and $\rho$ have independent dimensions. By applying the $\pi$ theorem from dimensional analysis (\citealt{10}), we obtain
\begin{equation}
	\frac{\Delta \sigma}{E}=\prod\left(\frac{w}{\rho d}, \frac{A_{i}}{d^{2}}, \frac{p}{d}, \frac{c}{d^{2}}, \frac{D}{d}, \frac{R}{d}\right).
	\label{eqn:3}
\end{equation}

In our final design, $w$, $d$, $R$, $D$, and $A_{i}$ are already determined; the weight of the reflector element $w$ is approximately $17 kg/m^{2}$, the diameter of the illuminated aperture $d$ is 300$m$, the radius of the cable-net $R$ is 300$m$, the diameter of the ring beam $D$ is 500$m$, and all cross sections of the cable $A_{i}$ have been determined according their loads. For a steel cable, $\rho$ is approximately $7850 kg/m^{3}$ . Thus, Eq.~(\ref{eqn:2}) can be simplified as
\begin{equation}
\frac{\Delta\sigma}{E}=\prod\left(\frac{p}{d}, \frac{c}{d^{2}}\right) \label{eqn:4}
\end{equation}

Furthermore, the nodes of the outer edge of the cable net are fixed to the ring beam and, in contrast to the case for the other cross nodes, their positions cannot be adjusted by the actuator. To expand the observation zenith angle as much as possible, the outer edge of the illuminated aperture should coincide with the basic spherical surface. Only then can the outer edge of the illuminated aperture arrive at the outer edge of the cable net. Under such a constraint, we can derive that
\begin{equation}
c=22500+2p\sqrt{67500}. \label{eqn:5}
\end{equation}

Equation (\ref{eqn:3}) can then be further simplified as
\begin{equation}
\frac{\Delta\sigma}{E}=\prod\left(\frac{p}{d}\right) \label{eqn:6}
\end{equation}

The elastic modulus of steel cable is about 200$GPa$, and thus, the only variable remaining in the implicit function of Eq.~(\ref{eqn:6}) is $p/d$ , which is the focal ratio of the telescope.

From Eq.~(\ref{eqn:6}) we know that the fatigue stress range of FAST cable is most dependent on the focal ratio of the telescope. A different focal ratio would lead to a different relative position between the illuminated aperture and the base plane, which is directly related to internal forces of the cable net and stroke of the actuator.

\begin{figure}[!htm]
\centering
\includegraphics[width=0.5\textwidth]{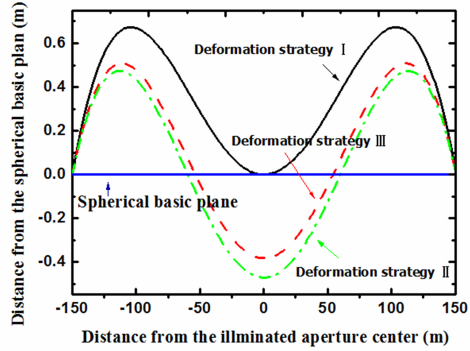}
\caption{Relative positions of the parabolic surface and the base spherical surface in three previously proposed deformation strategies}
\label{fig:3}
\end{figure}

In our earlier work (\citealt{6}), we proposed three deformation strategies, namely strategies I, II, and III. The relative positions of the illuminated aperture to the spherical base plane in these three strategies are shown in Figure.~\ref{fig:3}.

Among the strategies, strategies I and II respectively have the shortest actuator stroke and the minimum peak distance from the illuminated aperture to the base spherical surface. The actuator stroke is approximately 0.67$m$ in strategy I and the maximum deviation is approximately 0.47$m$ in strategy II. Strategy III is based on the principle of equal arclengths; i.e., the profile arclength of the illuminated aperture is equal to that of the spherical base plane. The focal ratios corresponding to these three strategies are respectively 0.4665, 0.4611, and 0.4613.

In the preliminary design of the FAST cable-net structure, deformation strategy I was recommended as the preferred control scheme simply because it has the shortest actuator stroke. Obviously, it is unreasonable to omit the fatigue problem in the optimization of the deformation strategy, especially in the present case where stress range is of the order of 500$MPa$.

The present work thus establishes a relation between the focal ratio and deformation stress range. By considering both the actuator stroke and stress range of the cable, we can reconsider our recommended deformation strategy for FAST observations.

\subsection{Simplified analysis method}

We use ANSYS software to build the finite element model of the entire supporting structure, which comprises cable net, down tied cable, ring beam, and steel pillar (see Fig.~\ref{fig:2}). Link10 elements and beam 188 elements are respectively used to simulate the response of the cable-net structure and the steel ring beam structure.

When not in operation, the FAST cable net should hold a spherical shape under the combined loads of gravity, initial stress, and the down-tied cable. To derive such a state, inverse iteration is applied. When FAST is making observations, the deformation procedure for forming the illuminated aperture from the spherical shape can be simulated by employing conventional iteration method.

According to the working principle of FAST, the motion of the cable net cross nodes during the form-changing operation is very slow. If we take the sphere center as the origin of coordinates and take the observation direction as the polar axis, the polar equation of the illuminated aperture can be expressed as
\begin{equation}
\sin^{2}\theta\cdot\rho^{2}-553.294\cdot\cos\theta\cdot\rho-166250=0 \label{eqn:7}
\end{equation}
where $\rho$ is the distance from the cross node to the sphere center and $\theta$ is the polar angle of the cross node (see Fig.~\ref{fig:1}).

In form-changing operation, the tangential displacement of the cross node is negligibly small. The velocity and acceleration of the cross node can then be expressed as
\begin{equation}
\left\{\begin{array}{l}
v(\theta)=\frac{d\rho}{d\theta}\cdot\frac{d\theta}{dt} \\
a(\theta)=\frac{d\left(\frac{d\rho}{d\theta}\cdot\frac{d\theta}{dt}\right)}{d\theta}\cdot\frac{d\theta}{dt}
\end{array}\right. \label{eqn:8}
\end{equation}

The tracking angular velocity of FAST is 15 degrees per hour. Substituting the angular velocity into Eq.~(\ref{eqn:8}), we find that the maximum speed of cross node is no more than 0.58$mm/s$. Thus, the form-changing operation of the FAST cable net is simplified as a quasi-static process here.

\begin{figure}[!htm]
\centering
\includegraphics[width=0.45\textwidth]{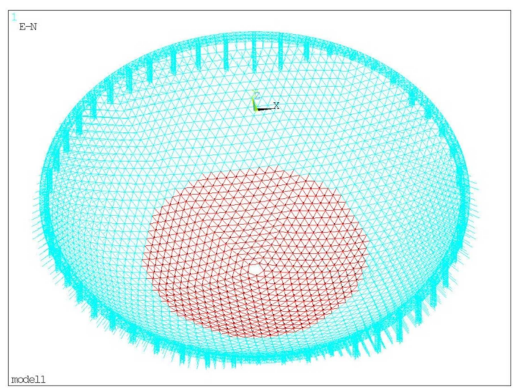}
\caption{Illustration of cross nodes used as discrete points to describe the continuous trajectory path of the illuminated aperture center}
\label{fig:4}
\end{figure}

According to the FAST working principle, the illuminated aperture center is restricted within a certain scope as shown in Fig.~\ref{fig:4}. This region contains 550 cross nodes, with the interval between any two adjacent nodes being no more than 12.5$m$, which corresponds approximately to a center angle of only 1 degree for the 300-$m$-radius reflector; this interval is much smaller than the illuminated aperture.

We assume that the distribution of the 550 discrete points is sufficiently dense, and any possible observation state is approximately equivalent to one of 550 deformation states whose illuminated aperture centers correspond to these 550 cross nodes. The peak stress range of each cable can then be easily derived from simulation results for the 550 deformation states.

To verify whether the distribution of discrete points is sufficiently dense, comparative analysis was performed for deformation strategy II using the abovementioned 550 discrete points and a denser distribution of discrete points. In the latter case, the discrete points include not only the cross-nodes but also the mid-point of each cable and the middle of each triangular element; thus, more than 2200 deformation states need to be simulated for the one deformation strategy.

\begin{figure}[!htm]
\centering
\includegraphics[width=0.8\textwidth]{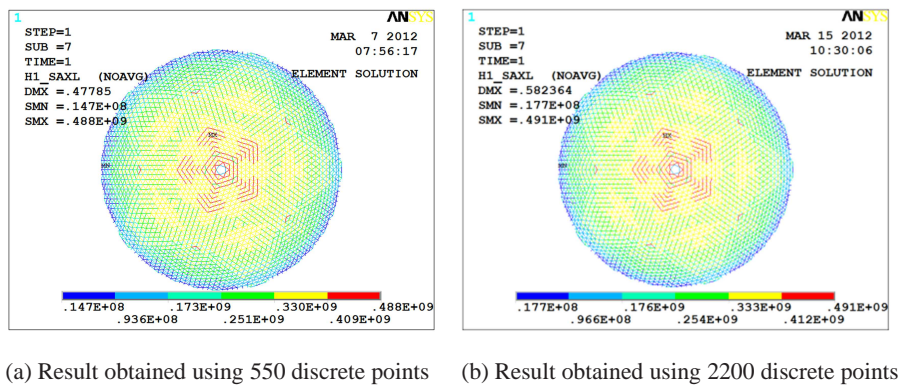}
\begin{tabular}{cc}
(a) Result obtained using 550 discrete points &
(b) Result obtained using 2200 discrete points
\end{tabular}

\caption{Comparison of the simulation results of deformation strategy II obtained using 550 and 2200 discrete points}
\label{fig:5}
\end{figure}

The calculation results derived using the two sets of discrete points described above are shown in Fig.~\ref{fig:5}a and b. The peak stress ranges derived using 550 and 2200 discrete points are respectively 488 and 491$MPa$, a difference of less than 1\%. We thus assume that using 550 cross nodes provides sufficient accuracy in our case. The subsequent work in this paper is based on this assumption.

\subsection{Analysis results}

Employing the same procedure used in the previous section, the peak stress range generated by several deformation strategies with different focal ratios can be derived (see Table~\ref{tab:1}). The peak stress range, taken as an analytical factor, is obtained for different focal ratios. The simulation results reveal that the stress range has a minimum value of 455$MPa$ when the focal ratio is 0.4621.

To verify that the optimum focal ratio is 0.4621 for the fatigue problem of the cable net, focal ratios of 0.4620 and 0.4622 were also investigated. The simulation showed that the peak stress ranges when employing these two deformation strategies are respectively 462 and 460$MPa$, both of them are slightly higher than that resulting when employing the deformation strategy with a focal ratio of 0.4621. Therefore, we have reason to believe that the deformation strategy with a focal ratio of 0.4621 is very close to the optimum deformation strategy having the minimum stress range.

It should also be noted that the actuator stroke in the strategy is 0.89m, which is 50$mm$ less than that in strategy II. By comparing the stress range and actuator stroke of the deformation strategies (see Table~\ref{tab:1}), we recommend a strategy with a focal ratio of 0.4621 for FAST application.

\begin{table}[!htm]
\bc
\begin{minipage}[]{100mm}
\caption[]{Comparison of the fatigue stress range of the cable and the stroke of the actuator when employing different deformation strategies\label{tab:1}}\end{minipage}
\setlength{\tabcolsep}{1pt}
\small
\begin{tabular}{ccccccccc}
	\hline\noalign{\smallskip}
	Focal ratio & 0.4603 & 0.4611 & 0.4613 & 0.4620 & \textbf{0.4621} & 0.4622 & 0.4633 & 0.4665\\
	\hline\noalign{\smallskip}
	Maximum stress range (MPa) & 512 & 488 & 482 & 462 & \textbf{455} & 460 & 488 & 547 \\
	\hline\noalign{\smallskip}
	Actuator stroke (m) & 0.9890 & 0.9450 & 0.9341 & 0.8966 & \textbf{0.8914} & 0.8861 & 0.8291 & 0.6741 \\
	\noalign{\smallskip}\hline
\end{tabular}
\ec
\end{table}

\section{Assessment of fatigue resistance}

According to the working principle of FAST, the problem of fatigue of the cable-net structure arises from the form-changing operation. The time history of the cable stress would directly depends on the trajectory of the illuminated aperture. Therefore, the present work on the assessment of fatigue of the cable net structure can be divided into several parts.

First, according to the scientific goal and observation model of the telescope, we roughly plan the trajectory of the illuminated aperture in the service life of the telescope of 30 years.

Second, employing a simplified finite element method, the stress–time history curve of each cable is derived from the trajectory of the illuminated aperture.

Finally, using a reasonable mathematical statistical method to deal with the stress–time history curve of the cable, we can count the approximate number of fatigue cycles of each cable.

\subsection{Planning the observation trajectory}

Regarding the scientific goals of FAST, observations made with the telescope can be classified into five classes: pulsars, neutral hydrogen, molecular spectral lines, very-long-baseline interferometry and the search for extraterrestrial intelligence. The observing mode can then be divided into three types: pulsar searching and monitoring, neutral-hydrogen large-area and small-area scanning, and other observations. We assume that each of the three types of observations accounts for one-third of the observation time.

According to unofficial statistics for the first half of 2009, the observation uptime of the Green Bank Radio Telescope is approximately 70\% to 80\%, and the uptime of the Xinjiang 25-meter radio telescope is approximately 74\%. Considering the system complexity in the FAST design, more maintenance time will be needed for FAST. It is thus reasonable to assume that the FAST uptime will be no more than 70\%.

\begin{figure}[!htm]
\centering
\includegraphics[width=0.6\textwidth]{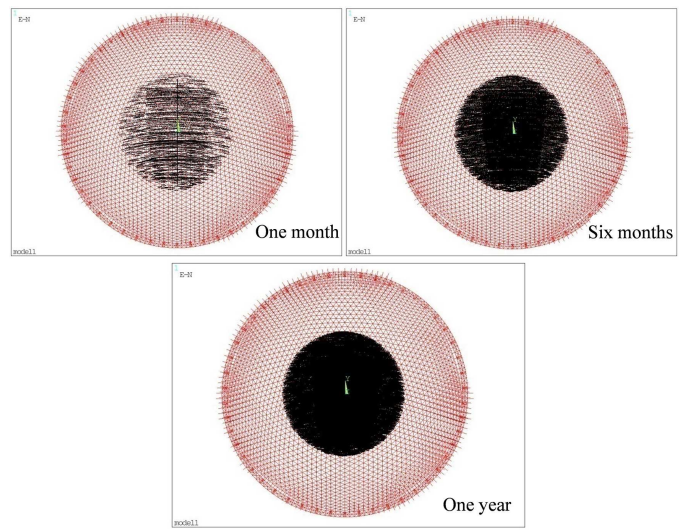}
\caption{Estimated trajectories of the parabolic center}
\label{fig:6}
\end{figure}

Making the above assumption and employing randomization principle, we can roughly plan the trajectory of the illuminated aperture during the service life of the telescope. A trajectory data file is then created. The data include a total of 228,715 observations and 3,410,008 tracking points. The interval time between tracking points is 120 seconds, and the corresponding spherical solid angle at the surface of the reflection is about 0.5 degrees. In employing the trajectory of the parabolic center to describe this problem, we can use MATLAB software to draw the observation
trajectory over different time periods, as shown in Figure~\ref{fig:6}.

According to the design principle of the reflector system, the edge of the cable net is fixed on the ring beam. The illuminated region cannot extend beyond the 500-meter diameter, and the maximum observation zenith angle is 26.4 degrees. The trajectory of the illuminated aperture is thus limited to within a certain region near the reflector center.

\subsection{Estimationof fatigue cycles}

It is unrealistic and unnecessary to use time history analysis to simulate the form-changing operation process. Fortunately, the form-changing operation of the FAST cable net can be simplified as a quasi-static process. We can thus simplify the continuous tracking process as a series of discrete deformation states.

We verified that any possible observation state is approximately equivalent to one of the 550 deformation states. Thus, with the above 3,410,008 tracking points, the stress–time curve of each cable can be estimated from the simulation results of the 550 deformation states.

Currently, the rain flow count method is the method most commonly used to analyze a fatigue stress spectrum (\citealt{11,12}). We can use this method to derive both the stress range and number of cycles. A program was thus written to apply the rain flow count method in the processing of the stress–time histories of all cables. The number of load cycles for each cable is thus derived.

The statistical results show that, in the abovementioned 228,715 observations, each cable went through between 840,107 and 1,020,054 cycles. The effect of the mean stress level is negligible when the mean stress is between 15\% and 40\% ultimate tensile strength (\citealt{13}). Our cable strength design fit well in this situation, so the effect of the mean stress is thus neglected here.

\begin{figure}[!htm]
\centering
\includegraphics[width=0.45\textwidth]{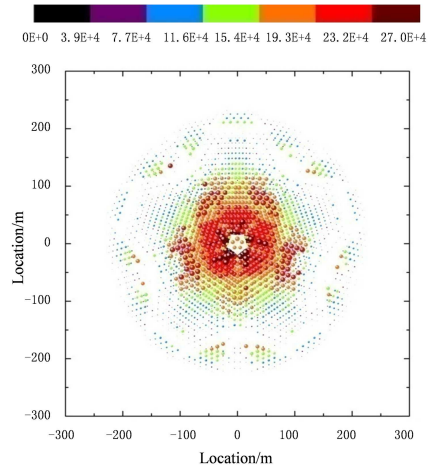}
\caption{Distribution of the stress cycles for a stress range exceeding 300$MPa$}
\label{fig:7}
\end{figure}

In general, certain materials have a fatigue limit or endurance limit that represents a stress level below which the material does not fail and can be cycled infinitely. Therefore, fatigue cycles when cables are in a higher stress range are of more concern here, especially when the stress range exceeds 300$MPa$. The distribution of the number of cycles when the stress range exceeds 300$MPa$ is plotted in Figure~\ref{fig:7}.

Figure~\ref{fig:7} shows that the stress range exceeds 300$MPa$ for 270,027 cycles. There is obvious regularity in that there are more fatigue cycles in the form-changing operation closer to the center of the reflector. It worth noting that the highest stress range and the maximum number of cycles are located near to one another.

\begin{figure}[!htm]
\centering
\includegraphics[width=0.5\textwidth]{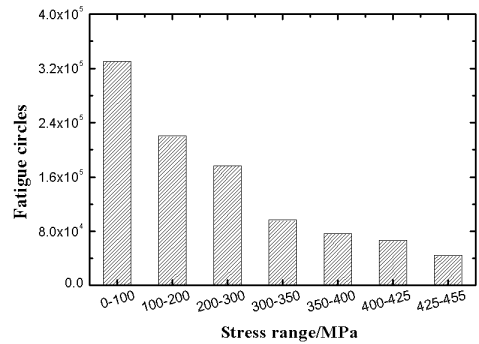}
\caption{Statistical result of the number of fatigue cycles versus the stress range}
\label{fig:8}
\end{figure}

By comparing the stress range and number of fatigue cycles of each cable, we can select a characteristic cable that we recognize as having a greater chance of failing to perform more detailed analysis. Figure~\ref{fig:8} shows the statistical results of the number of cycles versus the stress range in the abovementioned 228,715 observations.

There are $1.8\times 10^{5}$, $1.7\times 10^{5}$ and $1.1\times 10^{5}$ cycles for stress ranges of 200$\sim$300, 300$\sim$400 and 400$\sim$455$MPa$, respectively. There are approximately $4.6\times 10^{5}$ fatigue cycles for which the stress range exceeds 200$MPa$. Together with the stress range below 200$MPa$, the cycles number of this cable totally come to $1.02\times 10^{6}$ times.

\section{Fatigue experiment}

According to above results of numerical analysis, the cable stress range generated by shape-changing operation is about twice the standard authorized value. The designed service life of FAST is 30 years. However, in accordance with international practice, such a large radio telescope would in general have at least 50 years of service. Arecibo, for example, has been in service for more than 60 years. Meanwhile, the trajectory path of the illuminated aperture on the spherical cable-net is difficult to estimate accurately.

For reasons of security, the FAST project has proposed rigorous engineering requirements for the cable; i.e., the steel cable used in FAST construction should be able to endure $2\times10^{6}$ cycles of a fatigue test under a stress range of 500$MPa$. This is a serious challenge for the steel cable, with no successful experience for reference. Therefore, we start the present work from the basic tensile elements of the steel cable system; i.e., the steel wire and steel strand.

\subsection{Steel wire}

Previous research (\citealt{14}) has shown there is a direct relation between the fatigue resistance of a steel cable and that of its steel wire. It is necessary, especially in the case of the high fatigue resistance requirement of FAST, to start an investigation by focusing on the steel wire that is the base material of the steel cable system. The Post-Tensioning Institute has specified that the minimum fatigue test strength/performance of single steel wire is 370$MPa$ under $2\times10^{6}$ load cycles (\citealt{8}).

In general, the fatigue strength of a steel cable system, because of fretting fatigue and the non-uniform stress distribution, is somewhat lower than that of steel wire. Therefore, the abovementioned fatigue strength of a single steel wire would be unable to satisfy the requirement of FAST cable. Fortunately, improved materials have become available. We thus perform a fatigue test using the Chinese domestic supply of the latest high-performance steel wire.

\begin{figure}[!htm]
\centering
\includegraphics[width=0.5\textwidth]{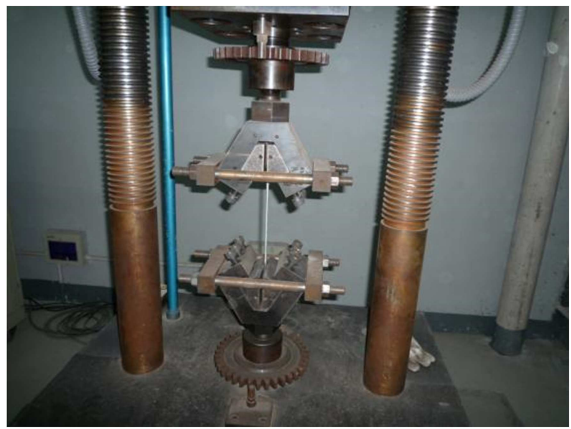}
\caption{Photograph of the fatigue test conducted on a single Super 82B galvanized steel wire}
\label{fig:9}
\end{figure}

Super 82B galvanized steel wire (1860-$MPa$ grade), manufactured by Baoshan Iron and Steel Company, Ltd., is selected to carry out the fatigue test. The tensile load fluctuates with a constant-amplitude sinusoidal shape, and the stress range is set at 600$MPa$ (from 144 to 744$MPa$). The purpose of the redundant 100$MPa$ is to allow for a reduction in the fatigue strength after the cable system is manufactured from the steel wires. The specimen is 200$mm$ in length and 5.20$mm$ in diameter(see Fig.~\ref{fig:9}). The experiment loading frequency is 10$Hz$.

The experimental results show that all six specimens endured $2\times 10^{6}$ cycles in the fatigue test. We thus conclude that the fatigue strength of a single steel wire has obviously improved beyond the historical value given above. However, it should be noted that fretting fatigue between adjacent wires will obviously affect the fatigue resistance. Further experimental investigation of the steel strand is still needed.

\subsection{Steel strand}

The problem of fretting fatigue is the most important factor affecting the fatigue resistance of a steel strand or cable system. We thus have reason to believe that the type of coating of steel wires will play an important role. Consequently, strands with different types of coating were tested in the present work, such as no coating, galvanized coating, and epoxy coating strands.

In general, epoxy-coated steel wire strands can be classified as filled epoxy-coated steel wire strands and individual epoxy-coated steel wire strands. The latter is selected in the present work mainly because its steel wires are individually isolated from each other by an epoxy coating, which efficiently reduces the stress concentration and friction damage on the surface of the steel wires.

All samples are in the form of a $1\times 7$ strand with nominal diameter of 15.2$mm$ and length of 1000$mm$. The stress range is conservatively set to 550$MPa$, which is 50$MPa$ higher than our requirement for the steel cable system.

\begin{figure}[!htm]
\centering
\includegraphics[width=0.4\textwidth]{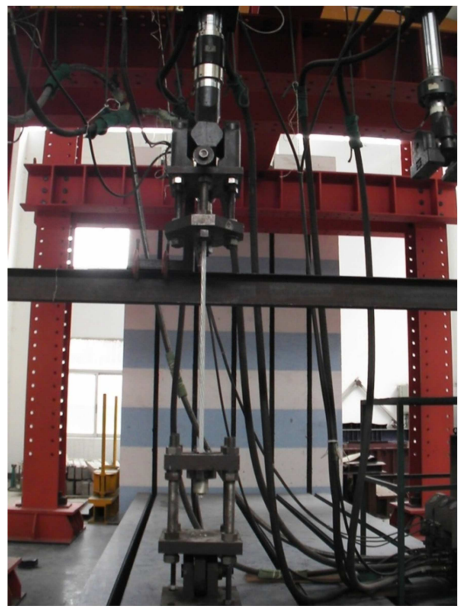}
\caption{Photograph of the fatigue test on a steel strand}
\label{fig:10}
\end{figure}

The tested strand was anchored by the wedge-type anchor, which is a common mechanism of anchorage. The fatigue test was performed on an MTS hydraulic fatigue machine frame (see Fig.~\ref{fig:10}). To eliminate the uneven distribution of stress, the specimen was subjected to initial loading from zero to about 80\% guaranteed ultimate tensile strength. Dynamic loading was then applied between 13.12\% and 40\% guaranteed ultimate tensile strength. The fatigue test was performed at a loading frequency of approximately 10$Hz$. The results for different strands are listed in Table~\ref{tab:2}.

\begin{table}[!htm]
\bc
\begin{minipage}[]{100mm}
\caption[]{Fatigue test results for different types of steel strand with a wedge-type anchor\label{tab:2}}\end{minipage}
\setlength{\tabcolsep}{2.5pt}
\small
\begin{tabular}{cccc}
	\hline\noalign{\smallskip}
	Coating & Load range & Stress range & Cycles\\
	\hline\noalign{\smallskip}
	No coating & 27.28-104.28 $kN$ & 550 $MPa$ & $3.0\times10^{5}$ \\
	No coating & 27.28-104.28 $kN$ & 550 $MPa$ & $2.88\times10^{5}$ \\
	No coating & 27.28-104.28 $kN$ & 550 $MPa$ & $2.07\times10^{5}$ \\
	No coating & 27.28-104.28 $kN$ & 550 $MPa$ & $2.8\times10^{5}$ \\
	No coating & 27.28-104.28 $kN$ & 550 $MPa$ & $1.5\times10^{5}$ \\
	Galvanized coating & 27.28-104.28 $kN$ & 550 $MPa$ & $4.56\times10^{5}$ \\
	Galvanized coating & 27.28-104.28 $kN$ & 550 $MPa$ & $5.58\times10^{5}$ \\
	Epoxy coating & 27.28-104.28 $kN$ & 550 $MPa$ & $2.0\times10^{6}$ \\
	Epoxy coating & 27.28-104.28 $kN$ & 550 $MPa$ & $2.0\times10^{6}$ \\
	Epoxy coating & 27.28-104.28 $kN$ & 550 $MPa$ & $2.0\times10^{6}$ \\
	\noalign{\smallskip}\hline
\end{tabular}
\ec
\end{table}

\begin{figure}[!htm]
\centering
\includegraphics[width=0.8\textwidth]{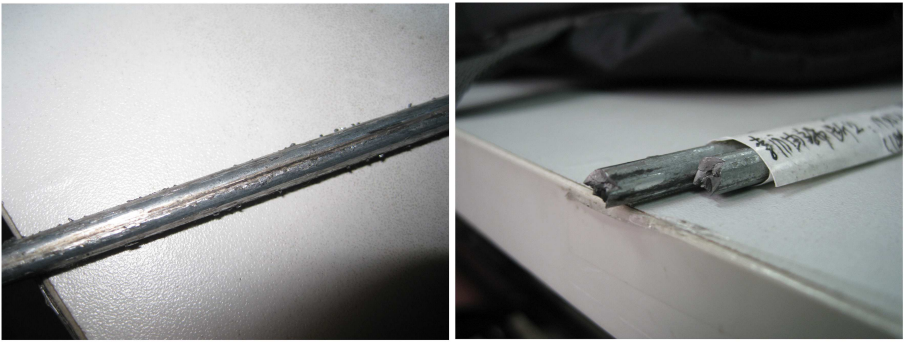}
\caption{Wear and scratching on the surface of a steel wire acquired from a steel strand broken in a fatigue test}
\label{fig:11}
\end{figure}

Table~\ref{tab:2} shows that all six strand samples with no coating broke within 300,000 cycles, and the two galvanized strands both broke at about 500,000 cycles. The present test result revealed that the obvious reduction in fatigue resistance of these two types of strands, compared with the fatigue resistance of their steel wires, may result from fretting fatigue between adjacent wires. We then took a steel wire from one broken strand sample to observe the friction at its surface. Figure~\ref{fig:11} shows an obvious scratch on the surface of this steel wire. Under repeated fatigue loading, such scratches are most likely to be the sources of initial cracks, and thus reduce the fatigue strength of the steel strand.

In the case of an individual epoxy-coated steel wire strand, the stress concentration and friction damage on the steel wire surface is efficiently reduced by the epoxy coating between the steel wires. Therefore, this type of strand was found to high fatigue resistance. In our test, all three samples of this type of strand endured $2\times10^{6}$ fatigue cycles under a stress range of 550$MPa$.

\subsection{S-N curve}

The cable-net structure is the most critical and expendable part of the FAST reflector system, and the service life of FAST directly depends on the residual fatigue life of the structure. FAST has a designed life of 30 years, and it is necessary to develop a fatigue damage monitoring system for such an expensive device.

The S-N curve is the basic data used in the evaluation of the structural fatigue life. However, it would be most unpractical and expensive to perform stay cable system fatigue tests at various numbers of load cycles and stress ranges to establish S-N curves. Fortunately, a significant
experience verified that the performance of an individual prestressing element, anchored with the actual anchorage details of the stay cable system, can be used as an indication of the approximate performance of the stay cable system (\citealt{8}).

Therefore, fatigue tests for different stress ranges were performed on epoxy coating steel strands Here, the upper stress was fixed at 744$MPa$, and different stress ranges of 550, 560, 570, 580, and 600$MPa$ were applied by changing the lower stress. Three samples were tested for each case.

\begin{figure}[!htm]
\centering
\includegraphics[width=0.5\textwidth]{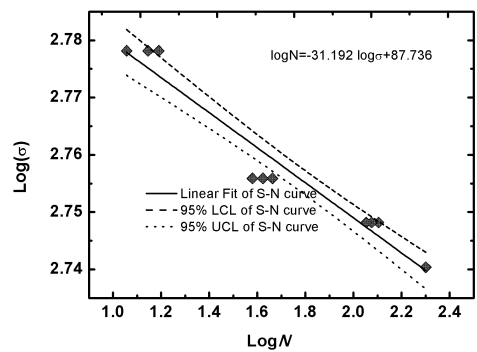}
\caption{Linear regression of Log ($\Delta\sigma$) versus Log ($N$) obtained in our experiment on the epoxy-coated strand}
\label{fig:12}
\end{figure}

The test results reveal that the fatigue life of a strand obviously declines with an increase in the stress range. At a stress range of 600$MPa$, strands generally broke after approximately 150,000 cycles. At a stress range of 580$MPa$, the three samples broke between 320,000 and 460,000 cycles. At a stress range of 560$MPa$, the three samples broke between 1,130,000 and 1,270,000 cycles. The relationship between the applied stress range and fatigue life can be plotted using dual logarithm coordinates as shown in Fig~\ref{fig:12}. Employing the least-squares method, the S-N curve of the type of strand investigated can be derived as
\begin{equation}
\log(N)=87.736-31.192\cdot\log(\Delta\sigma), \label{eqn:9}
\end{equation}
where $\Delta\sigma$ is stress range and $N$ is number of fatigue test cycles.

\subsection{Cable system}

For stay cable systems to achieve the specified minimum test performance requirements, the stay cable anchorage systems need to be carefully designed and detailed. In this work, traditional extruding anchoring technology was improved by adding a layer of cushioning material between the cable and anchoring device, and the anchoring system of the cable was realized by internal squeezing.

\begin{figure}[!htm]
\centering
\includegraphics[width=0.4\textwidth]{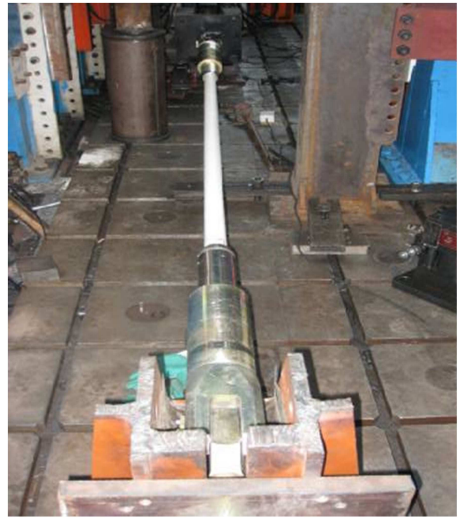}
\caption{Photograph of the location of the cable test}
\label{fig:13}
\end{figure}

Employing the improved anchoring, $3\times\Phi15.2$ and $6\times\Phi15.2$ cables, having effective cross-section areas of 420 and 840$mm^{2}$ respectively, were fabricated. The cross sections of tested cables are selected by referring to the actual cross-section selection used in FAST cable-net structure. Tests were carried out at the Supervision and Test Center for Product Quality belonging to the Ministry of Railways, and the Chinese Railway Bridge Bureau Group Corporation. Figure~\ref{fig:13} shows a tested cable.

According to the requirements of standards, the free segment length of the six cables was 3$m$. To eliminate the uneven distribution of stress, the specimen was subjected to initial loading from zero to about 80\% guaranteed ultimate tensile strength. The fatigue tests were performed under a fatigue stress amplitude of 500$MPa$, maximum stress of 744$MPa$ and loading frequency of 3$Hz$. The number of fatigue loadings reached 2 million without failure. Table~\ref{tab:3} gives the experimental results for the six cables.

\begin{table}[!htm]
\bc
\begin{minipage}[]{100mm}
\caption[]{Experimental results for cables\label{tab:3}}\end{minipage}
\setlength{\tabcolsep}{2.5pt}
\small
\begin{tabular}{cccc}
	\hline\noalign{\smallskip}
	Cable specifications & Stress amplitude	 & Number of cycles & Location \\
	\hline\noalign{\smallskip}
	$3\times\Phi15.2$ & 500 $MPa$ & 2 million times & Ministry of Railways \\
	$3\times\Phi15.2$ & 500 $MPa$ & 2 million times & Supervision and Test Center \\
	$3\times\Phi15.2$ & 500 $MPa$ & 2 million times & for Product Quality\\
	& & & \\
	$6\times\Phi15.2$ & 500 $MPa$ & 2 million times & Chinese Railway Bridge \\
	$6\times\Phi15.2$ & 500 $MPa$ & 2 million times & Bureau Group Corporation\\
	$6\times\Phi15.2$ & 500 $MPa$ & 2 million times & \\
	\noalign{\smallskip}\hline
\end{tabular}
\ec
\end{table}

\section{Conclusion}

During FAST observations, the stress range generated by the form-changing operation is more than twice the standard authorized value. To improve the reliability and service life of FAST, we carried out an extensive numerical and experimental investigation. The following conclusions are drawn from the results of the investigation.

\begin{enumerate}
\item The focal ratio is the key influencing factor of the stress range of the FAST cable-net structure generated by the shape-changing operation. Additionally, a focal ratio of 0.4621 is suggested as most appropriate, leading to a reduction in the stress range of approximately 30$MPa$ and a reduction in the actuator stroke of 50$mm$.

\item The tracking trajectory was planned according to the demands of the scientific objectives of FAST during its service life of 30 years. The technical requirements of the cables were obtained by mathematically simulating the tracking trajectory of FAST.

\item Compared with the historical value, there was an obvious improvement in the fatigue performance of a steel wire. In our fatigue test on the Super 82B steel wire, all six samples endured $2\times 10^{6}$ cycles under a stress range of 600$MPa$.

\item Because of fretting fatigue, different types of coating on the steel wire surface can obviously affect the fatigue performance of the strand. In our experiments on three types of coating strand, the best performance was found for the individual epoxy-coated steel wire strand, with all three samples enduring $2\times10^{6}$ fatigue cycles under a stress range of 550$MPa$.

\item S-N curves of the individual epoxy-coated steel wire strand were derived, giving basic data for the evaluation of the fatigue life of the FAST cable net in future operation.

\item The steel cable system was subjected to fatigue tests under a stress range of 500$MPa$ for 2 million fatigue loadings. We thus developed a steel cable system that could operate under high stress amplitude targeted at the relevant technical engineering requirements of FAST.
\end{enumerate}

\normalem
\begin{acknowledgements}
This work was supported by the Young Scientist Project of the Natural Science Foundation of China (Grant No. 11303059) and the Chinese Academy of Sciences Youth Innovation Promotion Association. We would like to thank all our colleagues for their contributions to our study.

\end{acknowledgements}
  
\bibliographystyle{raa}
\bibliography{paperef}

\end{document}